# Circular and Elliptic Restricted Three Body Problems -A Comparison

Robert Easton University of Colorado Department of Applied Mathematics
December 7, 2021

## Abstract:

The elliptic restricted three body problem has been well studied. However, the previous formulations of the problem have used a rotating coordinate system to keep the positions of the primary and secondary on the x-axis. This requires the rotation rate of the rotating coordinate system to speed up and slow down. The approach here uses a rotating coordinate system with constant rotation rate. This formulation allows one to conveniently compare solutions of the elliptic problem with solutions of the circular restricted three body problem.

## 1. Introduction:

Choosing a fixed Jacobi constant C for the restricted problem one can plot Hills boundaries and then choose initial positions and velocities as initial conditions for both the circular and the elliptic problems. One then plots numerical solutions of both sets of equations and plots their projections into configuration space for the planar and for the spacial problems. In this study we choose to compare the planar circular and elliptic problems.

## 2. Equations of Motion:

Capital letters are used for inertial coordinates. Start with the three body problem and set mass three to zero. The equations of motion for the restricted 3-body problem are:

$$\ddot{Q}_1 = Gm_2|Q_2 - Q_1|^{-3}(Q_2 - Q_1)$$

$$\ddot{Q}_2 = Gm_1|Q_2 - Q_1|^{-3}(Q_1 - Q_2)$$

$$\ddot{Q}_3 = Gm_1|Q_1 - Q_3|^{-3}(Q_1 - Q_3) + Gm_2|Q_2 - Q_3|^{-3}(Q_2 - Q_3)$$

By setting the center of mass $m_1 Q_1 + m_2 Q_2 = 0$, one can set the position of the secondary body $Q_2 = -\alpha Q_1, \alpha = m_1/m_2$. We have $Q_2 - Q_1 = -(1+\alpha)Q_1$. The resulting equations are:

(1) $\ddot{Q}_1 = -K|Q_1|^{-3}(Q_1), K = Gm_2(1+\alpha)^{-2}$

(2) $\ddot{Q}_3 = Gm_1|Q_1 - Q_3|^{-3}(Q_1 - Q_3) - Gm_2|\alpha Q_1 + Q_3|^{-3}(\alpha Q_1 + Q_3)$

To compare solutions of the elliptic problem with solutions of the circular restricted three body problem we choose a circular solution of (1) with radius $\rho$. The solution is $Q(t) = \rho e^{i\omega t}$ with $\omega^2 = K\rho^{-3}$.

A rotating coordinate system with this rotation rate will be used. In the rotating system the primary and secondary masses for the elliptic problem will trace ovals around the corresponding fixed positions of the primary and secondary masses for the circular problem. To quantify this study we take the mass of the primary to be the earth mass and the mass of the secondary to be the mass of the moon.

In order to compare solutions of the elliptic problem with the circular problem the equations (1) and (2) are transformed into a constantly rotating coordinate system. The transformation from a rotating frame to an inertial frame uses complex variable notation $Q_1 = e^{i\omega t}q, Q_3 = e^{i\omega t}x$.
If the motion in the inertial frame is modeled by the equation $\ddot{Q} = F(Q)$, and $Q = e^{i\omega t}q$, then $\ddot{Q} = e^{i\omega t}[\ddot{q} + 2i\omega\dot{q} - \omega^2 q] = F(Q)$, and the corresponding equation in the rotating frame is $\ddot{q} + 2i\omega\dot{q} - \omega^2 q = e^{-i\omega t}F(e^{i\omega t}q)$.

This formula is used to transform equations (1) and (2) into the rotating frame with rotation rate $\omega^2 = K\rho^{-3}$.

(3) $\ddot{q} + 2i\omega\dot{q} - \omega^2 q = -K|q|^{-3}q$

(4) $\ddot{x} + 2i\omega\dot{x} - \omega^2 x = Gm_1|q-x|^{-3}(q-x) - Gm_2|q-x|^{-3}(\alpha q + x)$

Then $q_0(t) = (\rho, 0)$ is an equilibrium solution of (3).

3. Equations for the Elliptic 3-Body Problem:

For the elliptic problem hoose an initial condition for equation (3) close to the equilibrium and then solve the combined system (3) and (4).

First change the distance scale to set the earth-moon distance $d$ to 1.
Set $dz = q$, $\mu = 1/(1+\alpha), \mu d = \rho$

$d(\ddot{z} + 2i\omega\dot{z} - \omega^2 z) = -Kd^{-3}|z|^{-3}dz$

Set $dX = x$,

$d(\ddot{X} + 2i\omega\dot{X} - \omega^2 X) = Gm_1 d^{-3}|z-X|^{-3}(dz - dX) - Gm_2 d^{-3}|z-X|^{-3}(\alpha dz + dX)$

Simplify

(5) $\ddot{z} + 2i\omega\dot{z} - \omega^2 z = Kd^{-3}|z|^{-3}z$

(6) $\ddot{X} + 2i\omega\dot{X} - \omega^2 X = Gm_1 d^{-3}|z-X|^{-3}(z-X) - Gm_2 d^{-3}|z-X|^{-3}(\alpha z + X)$

Now change the time scale to replace $d/dt$ by $\omega d/dt$ and replace $d^2/dt^2$ by $\omega^2 d^2/dt^2$

$$\omega^2 \ddot{z} + 2i\omega^2 \dot{z} - \omega^2 z = Kd^{-3}|z|^{-3}z$$

$$\omega^2 \ddot{X} + 2i\omega^2 \dot{X} - \omega^2 X = Gm_1 d^{-3}|z-X|^{-3}(z-X) - Gm_2 d^{-3}|z-X|^{-3}(\alpha z + X)$$

Simplify with $\omega^2 = K\rho^{-3}$, $Kd^{-3}/K\rho^{-3} = \mu^3$. Set $Gm_1/(d^3\omega^2) = \alpha\mu = 1-\mu$, $Gm_2/(d^3\omega^2) = \mu$. The final normalized equations for the elliptic problem are

(7) $\ddot{z} + 2i\dot{z} - z = -\mu^3|z|^{-3}z$

(8) $\ddot{X} + 2i\dot{X} - X = (1-\mu)|z-X|^{-3}(z-X) - \mu|\alpha z + X|^{-3}(\alpha z + X)$

---

### 3. Parameters and Initial Conditions:

Experimental parameters:

$Gm_1 = 3.986 \times 10^5 \; km^3/s^2$, $Gm_2 = 4.0903 \times 10^3 \; km^3/s^2$.
$d = 3.844 \times 10^5 \; km$ mean earth-moon distance
$e = .0549$ eccentricity of the moon
$\alpha = Gm_1/Gm_2 = 81.2977$
$\mu = 1/(1+\alpha) = 0.01215$
$\mu^* = (1-\mu)$
$\omega^2 = Gm_2\mu^2 d^{-3}$

The primary (Earth) travels on an elliptic orbit of period $2\pi$, and eccentricity $e$.
In the normalized inertial frame the equation is

(9) $\ddot{w} = \mu^{-3}|w|^{-3}w$.

We want to find the initial condition $[w(0), \dot{w}(0)]$ that produces an elliptic orbit of period $2\pi$ with elliptic parameters $(p, e)$ starting near the position $(-\mu, 0)$. The period condition requires that the semi-major axis of the ellipse is equal to $\mu$. This requires that
$(1/2)(p/(1+e) + p/(1-e)) = \mu$ and thus $p = \mu(1-e^2)$. We choose
$w(0) = (-p/(1+e), 0)$ as the periapsis point on the orbit. The parameters of the ellipse determine angular momentum $\sigma = (\mu^3 p)^{1/2} = w(0) \times \dot{w}(0)$. Set $\dot{w}(0) = (0, v)$ and solve for $v = \mu(1+e)(1-e^2)^{-1/2}$.

Transforming the elliptic solution $w(t)$ to rotating coordinates with $z(t) = e^{-it}w(t)$, we see that $z(0) = w(0)$ and $\dot{z}(0) = \dot{w}(0) - iw(0)$. The result is
$\dot{z}(0) = (0, \gamma); \gamma = \mu[(1+e)^{1/2}(1-e)^{-1/2} - (1-e)]$. This is the initial condition for equation (7). This models the motion of the primary (Earth) in the rotating normalized coordinate system.

The initial condition for (8) can be any condition of interest for the CR3BP. A computer code then computes and plots for comparison solutions for the ER3BP and the CR3BP starting from the same initial conditions.

## 4. Results:

Figure 1 compares solutions of the circular and elliptic problems starting at initial position $(.6,.2)$ with initial velocity direction $(1, -1)$ and Jacobi constant $C = 3.17$. The solutions track each other quite well.

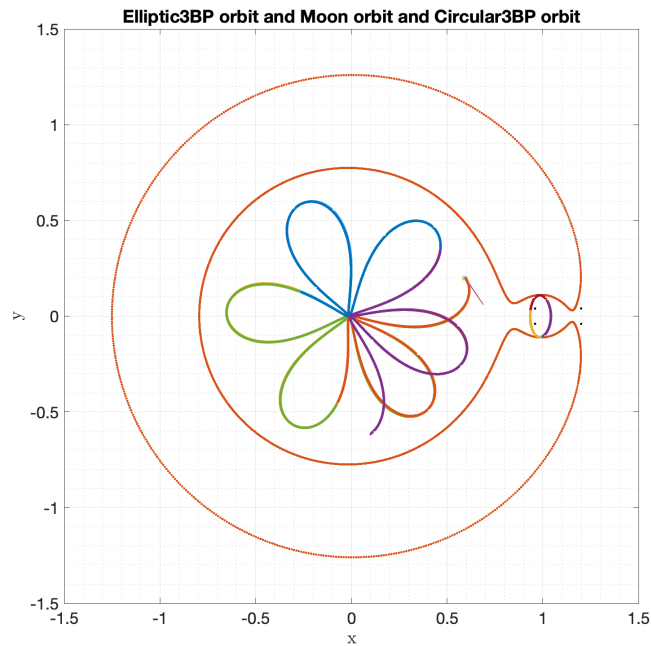

Figure 1 Hills Boundary, C = 3.17, Circular and Elliptic Orbits, and Lunar Orbit

Figure 2 compares solutions that diverge quickly. The orbit for the elliptic problem has a close encounter with the moon and crosses the Hills boundary. The orbit for the circular problem does not have a close flyby of the moon and remains within the Hills boundary.

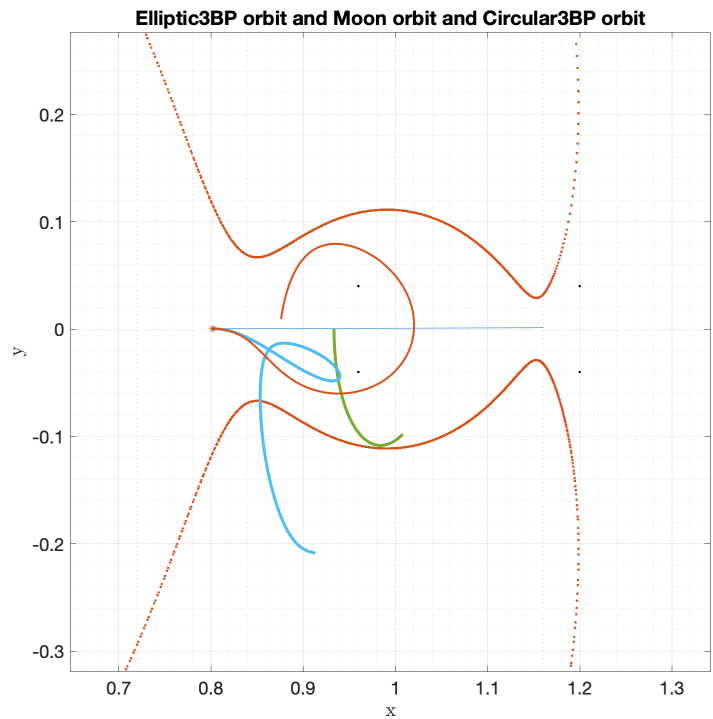

Figure 2 Orbits of the Circular and Elliptic problems, C = 3.17

## Conclusions:

The circular and elliptic restricted three-body problems have been studied extensively by many authors from Poincare and Hill up to the present time. The present article is a small contribution to this enormous literature and the reader is encouraged to do a google search for further references. The point of this article is to show that the two problems can be studied and compared using the same rotating coordinate system.

## Acknowledgements:

I would like to thank Martin Lo and Rodney Anderson for suggesting that I look at the elliptic problem, and for their collaboration over several years.